\documentclass[a4paper]{jpconf}

\usepackage[dvips]{graphicx}
\usepackage[]{amsmath}
\usepackage[]{amssymb}
\usepackage[]{latexsym}
\usepackage[]{endnotes}
\usepackage[]{bm}
\usepackage[]{wrapfig}

\newcommand{\mb}[1]{\mbox{\boldmath $#1$}}

\newcommand{\eas}[0]{\begin{eqnarray*}}
\newcommand{\eae}[0]{\end{eqnarray*}}
\newcommand{\les}[0]{\begin{equation}}
\newcommand{\lee}[0]{\end{equation}}
\newcommand{\leas}[0]{\begin{eqnarray}}
\newcommand{\leae}[0]{\end{eqnarray}}
\newcommand{\mchss}[4]
{
\left\{
\begin{array}{cc}
#1 & #2   \\
#3 & #4
\end{array}
\right.
}

\newcommand{\mmat}[4]
{
\left(
\begin{array}{cc}
#1 & #2 \\
#3 & #4 
\end{array}
\right)
}

\newcommand{\matthree}[9]
{
\left[
\begin{array}{ccc}
#1 & #2 & #3 \\
#4 & #5 & #6 \\
#7 & #8 & #9 
\end{array}
\right]
}

\newcommand{\mvec}[2]
{
\left(
\begin{array}{c}
#1  \\
#2  
\end{array}
\right)
}

\newcommand{\mvecthree}[3]
{\small
\left(
\begin{array}{c}
#1  \\
#2  \\
#3  
\end{array}
\right)
}

\newcommand{\mynote}[1]{}

\begin{document}
\title{Topological aspect of graphene physics}

\author{Y. Hatsugai}

\address{
Institute of Physics, University of Tsukuba, Tsukuba, 305-8571 Japan
}

\ead{y.hatsugai@gmail.com}

\begin{abstract}
Topological aspects of graphene are reviewed focusing on the
massless Dirac fermions with/without magnetic field.
Doubled Dirac cones of graphene are topologically  protected 
by the chiral symmetry.
The quantum Hall effect of the graphene is 
described by the Berry connection of a manybody state by the
 filled Landau levels
which naturally possesses non-Abelian gauge structures.
A generic principle of the topologically non trivial states as the  
bulk-edge correspondence is applied for graphene with/without magnetic
field and explain some of the characteristic boundary phenomena of graphene.
\footnote{Version:Aug.25 (2010)}
\end{abstract}

\section{Introduction}

Graphene as a playground of massless Dirac fermions
has a long history of theoretical study\cite{Wallace47,Lomer55,McClure56}.
 However its experimental realization is
 a real surprise since the massless Dirac fermion itself is 
topologically non trivial especially under a magnetic field\cite{Nov05,Zhang05}.
It implies that we have a chance 
to observe exotic
 topological phenomena in a real world, which have been 
 confirmed experimentally for the past few years. 

The quantized Hall conductance is one of the most well-known topological objects
as the Chern numbers of the Berry connection which 
describes the bulk (without boundaries).
It is also described by the edge states, supplemented by the Laughlin argument,
which gives a different topological number.
These two topological quantities
are closely related as the bulk-edge correspondence\cite{Hatsugai93b}.
This bulk-edge relation has been realized
 as a universal feature of topological states
in several different quantum systems such as superconductors, quantum 
(spin) Hall effects\cite{Qi10,Moore10,Hasan10}, 
gapped quantum magnets and photonic crystals\cite{Wang09}. 
The edge states in a magnetic field are chiral fermions in a sense, they have specific direction, 
and can not be destroyed as far as the bulk is gapped. It is
the topological stability of the edge states.
Graphene under a magnetic field is one of the typical topological states
where the bulk-edge correspondence of the Dirac fermions
plays a fundamental role.  
Localized zero modes near the 
zigzag boundaries (Fujita states)\cite{Fujita96},
which have been  measured by the STM experiments\cite{Kobayashi05},
and their generalization under a magnetic field,
are also governed by this topological principle\cite{Hatsugai93b}.
 Further appearance of two Dirac fermions with opposite chirality at 
the K and K' points are not accidental but is topologically 
protected by
the chiral symmetry which is a two dimensional analogue of the 
Nielsen-Ninomiya theorem in four dimensions\cite{Nielsen81}.

Also the Landau level of graphene is quite special
reflecting its singular dispersion as of the
Dirac fermions. Then the quantum Hall effects of it is anomalous\cite{Zheng02,Gus05,Hatsugai06gra}.
Further Landau degeneracy of the characteristic zero energy Landau level
are  topologically protected against for 
some class of randomness by the index theorem.
It is fundamental for 
 free-standing single layer graphene
where its intrinsic disorder is 
a gauge field fluctuation\cite{Guinea08,kawarabayashi09}.

Based on our works, 
these topological phenomena  
 in graphene physics will be explained intuitively.

\section{Topological stability of the Dirac cones and the $n=0$ Landau level}
\subsection{Chiral symmetry and doubled Dirac cones}
\begin{figure}[h]
\begin{center}
\begin{minipage}{6cm}
\begin{center}
\includegraphics[width=6.cm]{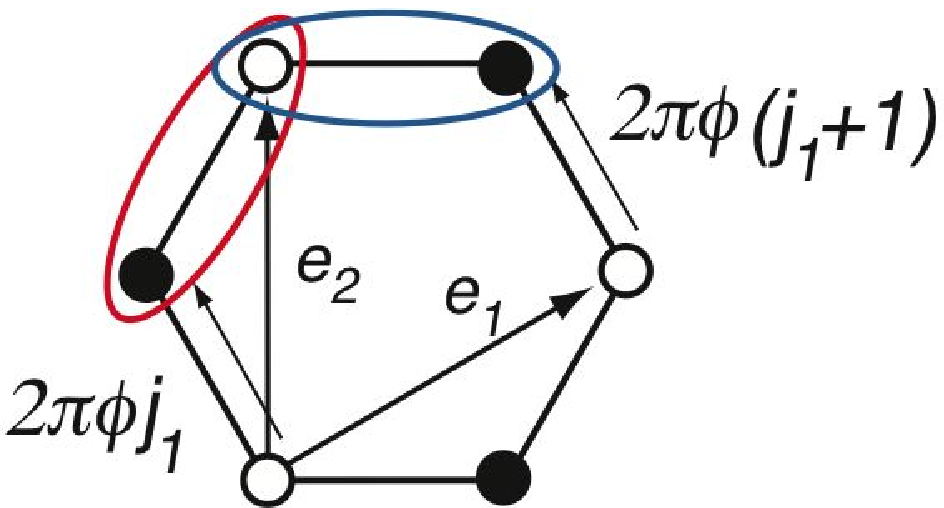}
\end{center}
\caption{\label{fig:unit}\small 
The unit cell of the graphene. The red ellipse is consistent with
 the zigzag edges and the blue one is for the bearded one.
Two primitive translation vectors are also shown. }
\end{minipage}\hspace{2pc}%
\begin{minipage}{7.5cm}
\begin{center}
\includegraphics[width=7.5cm]{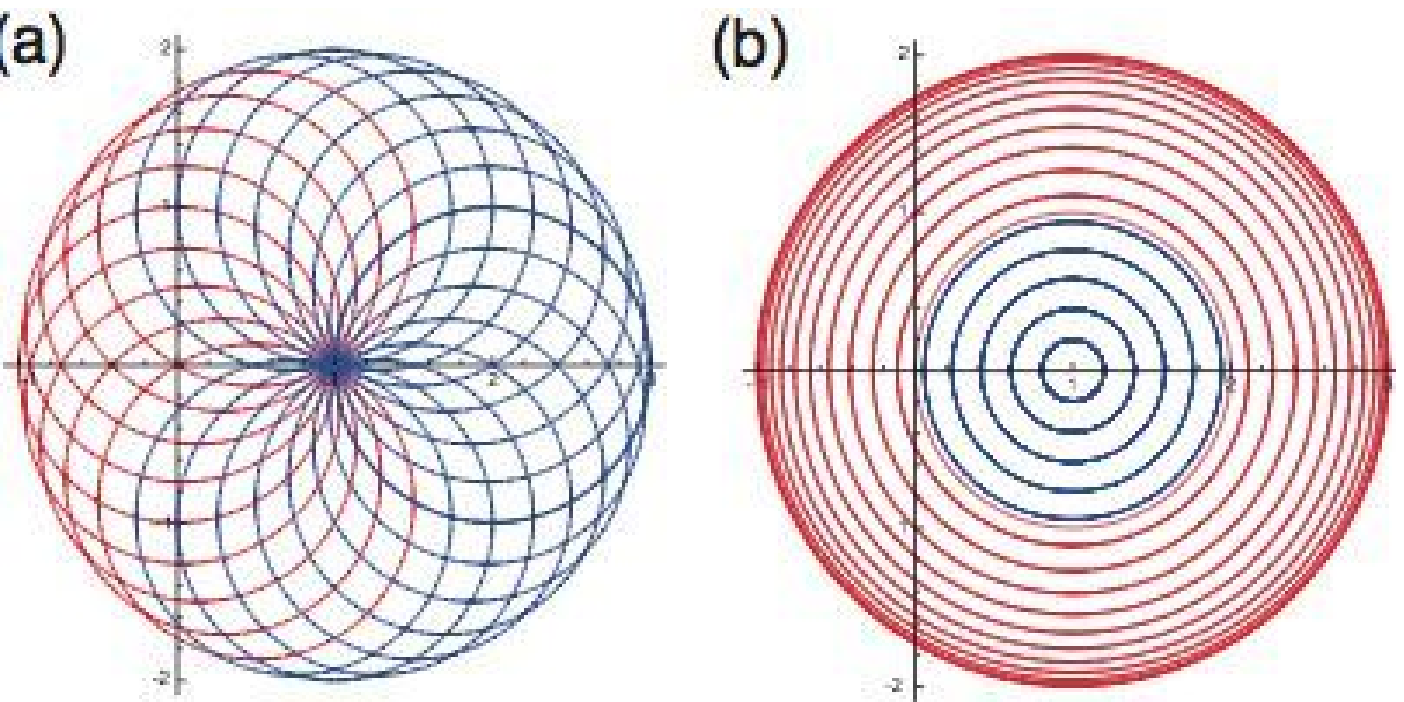}
\end{center}
\caption{\label{fig:DonC}\small Trajectories of 
$\Delta(\bm{k} ) $ parametrized by $k_1$ for various values of $k_2$.
 (a) for $\Delta =\Delta _Z $ and (b) $\Delta =\Delta _B$ ($t=-1$). Curves
that encloses the origin is drawn in red.}
\end{minipage} 
\end{center}
\end{figure}

Let us consider a single orbital tight-binding  hamiltonian of graphene 
with magnetic field\cite{Hatsugai06gra}
($ \bm{j}  =(j_1,j_2)$ is a two dimensional coordinate
with the unit translations $\bm{e}_{1,2} $ and 
$\phi$ is a total flux per hexagon in flux quantum $\Phi_0=h/e$,  See Fig.\ref{fig:unit})
\begin{eqnarray*}
H &=&
 t\sum_j\bigg[
 c_\bullet ^\dagger  ( \bm{j} ) c_\circ (\bm{j} ) 
+
 e^{2\pi \phi j_1 }
 c_\bullet ^\dagger  (\bm{j} ) c_\circ (\bm{j} -\bm{e} _2) 
+
 c_\bullet ^\dagger  (\bm{j} +\bm{e} _1) c_\circ (\bm{j} )  + h.c.
\bigg]\quad (\text{zigzag})
\\
&=&   t\sum_j\bigg[
 c_\bullet ^\dagger  ( \bm{j} ) c_\circ (\bm{j} ) 
+
 e^{2\pi \phi (j_1 +1)}
 c_\bullet ^\dagger  (\bm{j}) c_\circ (\bm{j}+ \bm{e}_1 -\bm{e}_2 )
+ c_\bullet ^\dagger  (\bm{j}-\bm{e}_1   ) c_\circ (\bm{j})
+ h.c.
\bigg]\quad (\text{bearded})
\end{eqnarray*} 
where 
we give two expressions using two different unit cells
that are compatible for zigzag and bearded edges.
When $\phi=0$, it is written in the momentum representation 
as 
$
H = \int \frac {d^2k}{(2\pi)^2} 
\bm{c}  ^\dagger (\bm{k} )
\bm{h} (\bm{k} )
\bm{c} (\bm{k} )
$,
\begin{eqnarray*}
\bm{h}_{Z,B} (\bm{k}  ) = 
\mmat
{0}
{\Delta_{Z,B} (\bm{k} )}
{\Delta_{Z,B} ^*(\bm{k} )}
{0},\quad
\mchss
{\Delta_Z(\bm{k} ) =t(1+e^{-ik_1}+e^{-ik_2})}{:(\text{zigzag})}
{\Delta_B(\bm{k} ) =t(1+e^{i(k_1-k_2)}+e^{ik_1})}{:(\text{bearded})}
\label{eq:Dham}
\end{eqnarray*} 
where
$\bm{c}  ^\dagger( \bm{k} )
=
 ( c_\circ ^\dagger( \bm{k} ) , c_\bullet ^\dagger( \bm{k} ) )$,
$
c_\alpha (\bm{j} ) = \int \frac {d^2k}{(2\pi)^2} e^{i \bm{k}\cdot  \bm{j} }
 c_\alpha (\bm{k} )
$.
The energy dispersion is given as $\epsilon(\bm{k} ) = \pm |\Delta (\bm{k} )|$ and
vanishing momenta $\bm{k}_D $, $\Delta(\bm{k}_D  )=0 $ give massless Dirac cones
if exist. 

Since the Brillouin zone, 
$T^2=\{(k_1,k_2)|k_1,k_2\in[0,2\pi]\}$, 
is a two-torus by identifying $k_1=0\leftrightarrows k_1=2\pi$
and $k_2=0\leftrightarrows k_2=2\pi$,
it does not have boundaries. Then considering $k_1$ as a
parameter, $\Delta (\bm{k}) $ makes a closed loop in the complex plane for each fixed $k_2$. This loop is modified 
by changing $k_2$ from $0$ to $2\pi$ and comes back to the original one
(Fig.\ref{fig:DonC}).
When the loop cuts the origin, it gives the Dirac cone.  
Then once the loop cuts the origin, the situation is stable
against for small but finite perturbation. This is the topological stability of the  Dirac cones. 
Here we need the $2\times 2$ hamiltonian is characterized by single complex parameter $\Delta $ 
( diagonal parts are zero). 
This is the {\em chiral symmetry}, that is, there exists 
$\{\bm{h} ,^\exists \bm{\gamma} \}=0$, $\bm{\gamma} ^2=\sigma _0$:
 $2\times 2$ unit matrix.  (In this case, $ \bm{\gamma }=\sigma _3$. )
Also the doubling of the Dirac cones is again clear 
since  the origin cutting the loop from the outside 
to the inside inevitably followed by cutting from inside to 
the out (See Fig.\ref{fig:DonC} and Fig.\ref{fig:doubling}).
 This simple observation
guarantees that the number of Dirac cones are always even. 
Note here that the chiral symmetry is related to the bipartite structure
of the honeycomb lattice, that is, the hoppings are only between the
two sub-lattices  $\circ $ and $\bullet $

\begin{figure}[h]
\begin{center}
\includegraphics[width=15.1cm]{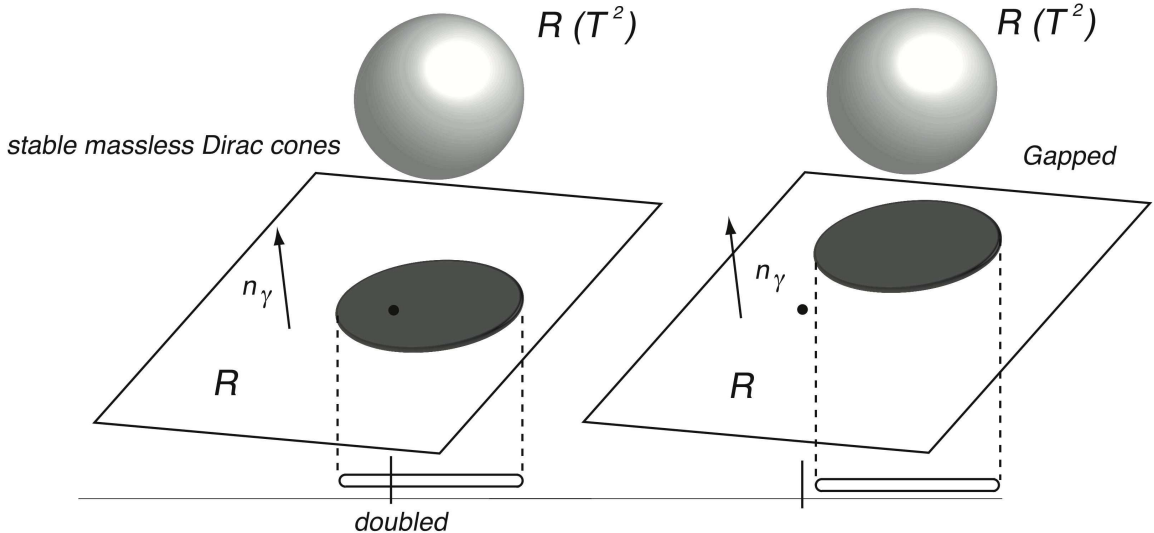}\hspace{2pc}%
\begin{minipage}[b]{16.0cm}
\caption{\label{fig:collapsed}\small 
Topological stabilities of the Dirac cones.
The image of the Brilloine zone $\bm{R}(T^2 )  $ is generically  a
surface in three dimensional space $\bm{R}  $, which is collapsed into
the plane which is normal to the $\bm{n}_\gamma  $, when
the hamiltonian is chiral symmetric as $\{H,\gamma \}=0$,  $\gamma =\bm{\sigma }\cdot \bm{n}_\gamma   $. 
}
\end{minipage}
\end{center}
\end{figure}

On the other hand, as for generic semiconductors including graphene,
let us assume  the effective hamiltonian of the valence and the 
conduction bands, $\bm{h} $, is chiral symmetric, 
that is, it does anti-commutes with 
some chiral matrix $\bm{\gamma } $, $\{\bm{h},\bm{\gamma }\}=0  $. 
It implies $\bm{h} $ is traceless and thus one can expand it 
by the Pauli matrices as $\bm{h}(\bm{k} )=\bm{R}(\bm{k} )\cdot \bm{\sigma }$
with three real parameters $R_1(\bm{k} )$, $R_2(\bm{k} )$ and $R_3(\bm{k} )$
 that makes a vector
 ${\bm{R}}(\bm{k} )=\mvecthree{R_1(\bm{k} )}{R_2(\bm{k} )}{R_3(\bm{k} )} $. 
The momentum dependent energy gap is 
$E_g(\bm{k})=2| \bm{R}(\bm{k} ) |$.
Then the zero gap condition is $\bm{R}=\bm{0}$.
 Writing the chiral matrix as,
 $\bm{\gamma }=\bm{n}_\gamma \cdot \bm{\sigma }   $ ($\bm{n}_\gamma ^2=1 $), the chiral symmetric condition
$\{\bm{h},\bm{\gamma }\} =
(\bm{R}\cdot \bm{n}_\gamma ) \sigma _0+i (\bm{R} \times \bm{n}_\gamma ) 
\cdot \bm{\sigma } +(\bm{R}\rightleftarrows \bm{n}_\gamma   )
= 2 (\bm{R}\cdot \bm{n}_\gamma ) \sigma_0 =0$,
implies $\bm{R}\cdot \bm{n}_\gamma =0  $, that is, $\bm{R} $ is 
always in the plane $\mathbb{R}^2(\bm{n}_\gamma )  $ whose normal vector is 
$\bm{n}_\gamma  $. Since the Brilluine zone $T^2$ is a 2-dimensional closed surface, the image
$\bm{R} (T^2 )$ is also a closed surface (like a balloon) in 3-dimensions. 
Then the condition 
$\bm{R}(T^2)\subset \mathbb{R}^2(\bm{n}_\gamma ) $ 
means the closed surface is collapsed on the plane (a rubber balloon without air
on the desk). 
When the collapsed image includes the origin, it gives the Dirac cones.
Then the topological stability of the Dirac cones and the doubling of them
are clear\cite{Hatsugai06gra,Hatsugai07gra,Hatsugai10}. 
(See Figs\ref{fig:collapsed}.)
It is the  2-dimensional analogue of the
Nielsen-Ninomiya theorem in the 4-dimensional lattice gauge theory
\cite{Nielsen81}. 
A relation between the  4D graphene and chiral fermions
is recently discussed as well\cite{Creutz08}.

\subsection{Chiral symmetry and the $n=0$ Landau level}
Now let us expand  the hamiltonian
 $\bm{h}(\bm{k} ) $ with chiral symmetry 
($\{\bm{h},\bm{\gamma }  \}$)
 near one of the doubled zero gap momentum, say, $\bm{k}_D$.
Then the effective hamiltonian in the lowest order in
 $\delta \bm{k}=\bm{k}-\bm{k}_D  $ is written as
$
\bm{h}  \approx   \bm{h} _C = 
\delta k_1 \bm{\sigma }\cdot \bm{X}
+
\delta k_2 \bm{\sigma }\cdot \bm{Y}
$ 
where 
$\bm{X} =     \partial _{k_1} \bm{R}$ and
$ \bm{Y} =    \partial _{k_2} \bm{R}$
which are perpendicular to the $\bm{n}_\gamma$.
It implies the chiral symmetry of the effective hamiltonian,  
$\{\bm{h}_C ,\gamma\}=0$.
One may define the chirality of the
effective hamiltonian $\chi_D=\pm 1$ as
$ \bm{n}_\gamma   = \chi_D {\bm{X}\times \bm{Y}  }/(c\hbar)^2$
where the effective {\em light velocity } ($c>0$) is defined
as $ c^2 \equiv|\bm{X}\times \bm{Y}  |/\hbar ^2$.
When the vector, $(\bm{X}, \bm{Y}, \bm{n}_\gamma )   $, forms
a right handed triple, the chirality, $\chi_D$ is +1 and, if left-handed, 
$\chi_D=-1$. Then the  chiral matrix is written as
$
\bm{\gamma } = \chi_D \bm{\sigma }\cdot( \bm{X}\times \bm{Y}) /(c\hbar)^2
$.

By the inverse procedure of the quantization,
$\hbar \delta \bm{k}\to
\bm{p} -e \bm{A} \equiv \bm{\pi} $, 
$p_\alpha = -i\hbar  \partial_\alpha$,  ($\alpha =x,y$),
we have a real space form of the effective hamiltonian 
with  a magnetic field,
$B= \partial _x A_y-\partial_y A_x  >0 $, $[\pi_x,\pi_y]=i\hbar eB$ as 
\begin{eqnarray*}
\bm{h}_C  &=&
\hbar ^{-1}\big[
  (\bm{\sigma } \cdot \bm{X} ) \pi_x
 + (\bm{\sigma } \cdot \bm{Y} ) \pi_y
\big]
\end{eqnarray*} 

The spectrum and the wave functions 
are determined
by considering its square
as 
\begin{eqnarray*} 
\bm{h} _C^2 &=& 
\hbar ^{-2} \big[ \bm{X} ^2 \pi_x^2+\bm{Y} ^2 \pi_y^2
+ (\bm{X}\cdot \bm{Y}  )
(\pi_x\pi_y
+
\pi_y\pi_x
)+ i (\bm{X}\times \bm{Y}  )\cdot \bm{\sigma }
[\pi_x,\pi_y]\big]
\\
&=&  c^2\, ( \bm{\pi} ^\dagger 
\bm{\Xi}\,  \bm{\pi} )\bm{\sigma} _0 
-\chi_D (eB\hbar c^2)   \bm{\gamma}
\end{eqnarray*}
where
\begin{eqnarray*}
\bm{\Xi} &=&
 \frac {1}{(\hbar c)^2} (\bm{X},\bm{Y}  )  ^\dagger   (\bm{X} ,\bm{Y} )=
 \frac {1}{(\hbar c)^2} \mmat
{\bm{X}\cdot \bm{X}  }{\bm{X}\cdot \bm{Y}  }
{\bm{X}\cdot \bm{Y}  }{\bm{Y}\cdot \bm{Y}  }
\\
\det \bm{\Xi} &=& \frac {1}{(\hbar c)^4}
\big( |\bm{X}|^2|\bm{Y}|^2-(\bm{X}\cdot \bm{Y})^2    \big)
=  \frac {|\bm{X}\times \bm{Y}|^2  }{(\hbar c)^4}=1
\end{eqnarray*} 

By identifying as  $\frac {1}{2m}\equiv c^2 $,
the first term of $\bm{h}_C $
is 
considered as a hamiltonian of the standard parabolic electrons 
with anisotropic masses (See \ref{sec:app1}).
Here the cyclotron frequency $\omega_C$ is identified
as $\omega _C=eB/m= 2eBc^2$.
Also note that the chiral matrix, 
 $\bm{\gamma} $,
commutes with the squared hamiltonian $\bm{h} _C^2$. 

Then writing 
 a normalized standard Landau state 
with energy, $\hbar \omega_C (n+1/2)=2\hbar eB c^2(n+1/2)$, $n=0,1,2,\cdots$,
as $\psi_n (x,y)$, 
the eigen state of $\bm{h}_C^2 $
 is written as
$\bm{\Psi}_n^\pm=\bm{\varphi}_\pm\psi_n(x,y)  $ where $\bm{\varphi}_\pm $
 is an eigen state of $\bm{\gamma } $ with eigen values $\pm 1$ respectively.
It satisfies
$
\bm{h} _C^2 \bm{\Psi}_n^\pm
= (eB\hbar c^2)(2n+1\mp\chi_D)\bm{\Psi}_n^\pm
$.
It implies that the effective hamiltonian $\bm{h}_C $ always 
has a zero energy Landau level as
\begin{eqnarray*}
\bm{\Psi}^0_{\chi_D}(x,y)\equiv \bm{\varphi}_{\chi_D}\psi_0(x,y) &:& 
\bm{h}_C \bm{\Psi}^0_{\chi_D}(x,y) =0
\end{eqnarray*} 
which has the chirality $\chi_D$. 

As for the non zero energy Landau levels, $\bm{h}_C^2 $ is doubly 
degenerated at $\epsilon_n^2$,  $\epsilon_{n} =c\sqrt{2neB\hbar} $, 
spanned by
$
 \bm{\Psi}_n^{+\chi_D}=\bm{\varphi}_{+\chi_D} \psi_n(x,y)$ and
$
 \bm{\Psi}_{n-1}^{-\chi_D}=\bm{\varphi}_{-\chi_D} \psi_{n-1}(x,y)
$, 
which diagonalize $\bm{h}_C $ without degeneracy
with the energies $\pm \epsilon_{n} $. 

\subsection{Chiral symmetry and Fermion doubling}
Up to this point, 
we have discussed the effective theory of the chiral symmetric 
zero gap semiconductor
near one of the gapless momentum. 

\begin{wrapfigure}{r}{6cm}
\begin{center}
\includegraphics[width=7.0cm]{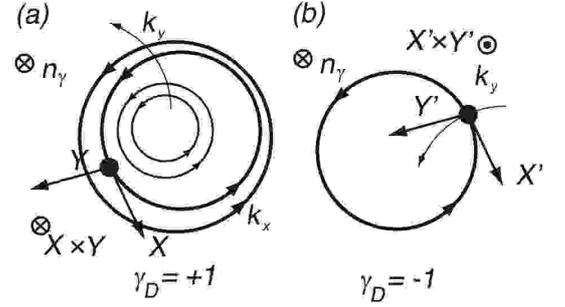}
\caption{\label{fig:doubling}\small 
$\bm{R}(\bm{k} )  $ on the plane $\mathbb{R}^2(\bm{n}_\gamma )$.
 (a) When the curve cuts the origin which defines the  Dirac cone.
(b) Another Dirac cone which is paired with the one in (a). The chirality is reversed.
}
\end{center}
\end{wrapfigure}
As discussed the chiral symmetry of the global hamiltonian $\bm{h} $,
$\{\bm{h}(\bm{k}),\bm{\gamma }\}=0   $, ($\bm{h}= \bm{R}\cdot \bm{\sigma } $ )
requires that the three dimensional vector
$\bm{R}(^\forall\bm{k} ) $ is always perpendicular 
to the $\bm{n}_\gamma  $ and on the plane 
$\mathbb{R}^2(\bm{n}_\gamma ) $
cutting the origin $\bm{R}=0 $
( $\bm{R}(\bm{k})\in \mathbb{R}^2(\bm{n}_\gamma )   $).
Then one can repeat the discussion for $\Delta(\bm{k} ) $
in Fig.\ref{fig:DonC} for the contour $\bm{R}(\bm{} ) $ on 
the plane $\mathbb{R}^2(\bm{n}_\gamma  )$, that is, 
considering a closed curve characterized by $k_2$, 
$C(k_2)=\{\bm{k}|k_1\in [0,2\pi]\} $. This curve is on the plane
$\mathbb{R}^2(\bm{n}_\gamma ) $ and comes back to the original one 
by changing $k_1$ form $0$ to $2\pi$. 
When it cuts the origin, it gives the Dirac cones. 
The doubling of the Dirac cones is also clear as before
 (Figs.\ref{fig:DonC} and \ref{fig:doubling}).
Also as for the chirality of the Dirac cones, $\chi_D$,
the paired Dirac cones, $D$ and $D'$, have a reversed chirality
as
\begin{eqnarray*}
\chi_D +\chi_{D'} &=& 0
\end{eqnarray*}
which is also clear from the Fig.\ref{fig:doubling}.
Then the chirality of the $n=0$ Landau level is also 
reversed. 
As for the graphene, the Dirac fermions at K and K' do have 
the reversed chirality.

\subsection{Aharonov-Casher argument}
As discussed the $n=0$ Landau level is an eigen state
of the chiral operator for each Dirac fermion
which is a characteristic feature of the chiral symmetric
massless Dirac fermions on lattice. It is realized in graphene. 
This complete  degeneracy of the Landau level is
not only for the pure system and it does persist for some class
on the randomness\cite{Kawarabayashi09}. There is well-known 
discussion by Aharonov-Casher\cite{Aharonov79} which is considered as 
a direct demonstration of the index theorem
where the degeneracy is only determined by the total flux passing through the
system as far as the effective description by the Dirac fermions is allowed. 
Here the magnetic field $B$ may not be uniform. 
We compactly describe it  extending the discussion to our anisotropic case.
 Since the anisotropic mass matrix $ \Xi  $ discussed before
is real symmetric and its determinant unity,
 it is diagonalized by the orthogonal matrix  $\bm{V} $ as (See also appendix),
$
\mb{\Xi}  = \mb{V} ^\dagger  \mmat{\xi_X}{}{}{\xi_Y} \mb{V} 
$, 
where $\xi_X,\xi_Y>0$,  $\xi_X\xi_Y=1$.
Then defining canonical momenta $\bm{\Pi}=\bm{V}\bm{\pi}   $, 
$[\Pi_X,\Pi_Y]=[\pi_x,\pi_y]=i\hbar eB]$, $\bm{h} _C^2$ is factorized as
\begin{eqnarray*}
\bm{h} _C^2 &=& 
c^2(\xi_X\Pi_X^2+\xi_Y\Pi_Y^2)+ i\chi_D \gamma [\Pi_X,\Pi_Y]) 
= 
c^2 {\cal D } ^\dagger  {\cal D }
\\
{\cal D} &=& \sqrt{\xi}_X\Pi_X +i \chi_D\gamma \sqrt{\xi}_Y\Pi_Y
\end{eqnarray*} 
Then writing the vector potential as 
$
{A}_X= -\xi_Y\partial _Y\phi
$,
$
{A}_Y= \xi_X\partial _X\phi
$
and defining 
$\tilde X=X/\sqrt{\xi}_X$ and $\tilde Y=Y/\sqrt{\xi}_Y$
(
$\sqrt{\xi}_X \partial _X= \partial_{\tilde X}$
and 
$\sqrt{\xi}_Y \partial _Y= \partial_{\tilde Y}$
),
the zero mode $\Psi^0_{\chi_D} =\varphi_{\chi_D}\psi$ satisfies, 
\mynote{
\begin{eqnarray*}
0 &=& {\cal D} \psi^0 _{\chi_D}
=
\bigg[
\sqrt{\xi}_X\big(-i\hbar \partial _X+ e \xi_Y\partial _Y\phi\big)
+i \chi_D^2
\sqrt{\xi}_Y\big(-i\hbar \partial _Y- e \xi_X\partial _X\phi\big)
\bigg]\psi^0_{\chi_D}
\\
&=& 
\bigg[
-i\hbar \sqrt{\xi}_X\partial _X+ e \sqrt{\xi}_Y\partial _Y\phi
+
\hbar \sqrt{\xi}_Y \partial _Y-i 
e\sqrt{\xi}_X \partial _X\phi
\bigg]\psi^0_{\chi_D}
\\
&=& 
\bigg[
-i\hbar \partial _{\tilde X}+ e \partial _{\tilde Y}\phi
+
\hbar \partial _{\tilde Y}-i 
e\partial _{\tilde X}\phi
\bigg]\psi^0_{\chi_D}
\\
&=& 
-i\hbar 
\bigg[
\partial _{\tilde X}+i \frac {2\pi}{\phi_0}  \partial _{\tilde Y}\phi
+ i
\partial _{\tilde Y}+
 \frac {2\pi}{\phi_0}  \partial _{\tilde X}\phi
\bigg]\psi^0_{\chi_D}
\end{eqnarray*}
}
$
0 = {\cal D} \psi^0 _{\chi_D}
= 
-i\hbar 
\bigg[
\partial _{\tilde X}+i \frac {2\pi}{\phi_0}  \partial _{\tilde Y}\phi
+ i
\partial _{\tilde Y}+
 \frac {2\pi}{\phi_0}  \partial _{\tilde X}\phi
\bigg]\psi^0_{\chi_D}
$.
Then writing 
as $
\psi_0 = e^{-2\pi 
 \frac {\phi}{\phi_0} }f
$,
the function $f$ satisfies
$
(
\partial _{\tilde X} + i
 \partial _{\tilde Y}
)
f = 0
$.
This is written by defining $z=\tilde X+i \tilde Y$ as
\mynote{
($\tilde X = (z+\bar z)/2$ and
$\tilde Y = 
 (z-\bar z)/2i$
 )
\begin{eqnarray*}
\partial _{\bar z} &=& \frac {\partial  }{\partial \bar z } 
=
\frac {\partial \tilde X }{\partial \bar z } \partial _{\tilde X}
+
\frac {\partial \tilde Y }{\partial \bar z } \partial _{\tilde X}
=\frac {1}{2} ( \partial _{\tilde X}+ i 
\partial _{\tilde Y})
\end{eqnarray*} 
}
$
\partial _{\bar z} f = 0
$.
Thus the function $f$ is an entire function of $z$ in the whole complex plane $z\in \mathbb{C}$, 
that is, polynomials. 
Also note that the function $\phi$ needs to satisfy 
$
B = \partial _X A_Y-\partial _Y A_X
= (\xi_X \partial _X^2+\xi_Y\partial _Y^2 )\phi
= ( \partial _{\tilde X}^2+\partial _{\tilde Y}^2 )\phi
$.
It implies
$ 
\phi(\tilde X,\tilde Y) = \int d\tilde Xd\tilde Y 
G(\tilde X-\tilde X',
\tilde Y-\tilde Y') B(\tilde X',\tilde Y')
$ 
where
$ G(\tilde X,\tilde Y) = 
 \frac {1}{2\pi} \log \frac {r}{r_0}
$, $ r^2=\tilde X^2+\tilde Y^2$ and $r_0$ is a constant. 
When assuming the magnetic field is only nonzero in the finite region, we have
for the asymptotic behavior in the limit $r\to \infty$ as 
$
\phi \stackrel{r\to\infty}{\longrightarrow} \frac {\Phi}{2\pi} \log\big(\frac {r}{r_0} \big)
$
where
$
\Phi = \int d\tilde Xd\tilde Y  \,B
=\int dX dY \, B
$
is a total flux.
Then we have for the asymptotic behavior of the
zero mode as
$
\psi \stackrel{r\to\infty}{\longrightarrow} f(z)
\big(\frac {r}{r_0} \big)^{-
 \frac {\Phi}{\phi_0} }
$. 
It implies 
that  the number of the degeneracy is $\frac {\Phi}{\phi_0} $.

\section{Hall conductance and Berry connection of Dirac sea}
Geometrical phase of the quantum mechanical states
are of fundamental interest for physical society today.
The quantum Hall effect and its  time reversal invariant analogue 
as
the quantum spin Hall effect are  the typical examples. 

Now let us start from the Niu-Thouless-Wu formula for the Hall conductance\cite{Niu85}.
It reads that 
the Hall conductance of the manybody state is given by the average over the
 twisted boundary condition as
\begin{eqnarray*}
\sigma_{xy}  &=& 
 \frac {e^2}{h} C,\ \
C =   \frac {1}{2\pi i} \int_0^{2\pi} d\phi^1\int_0^{2\pi} d\phi^2\, 
\big[
\langle \partial _1 G| \partial _2 G \rangle 
-
\langle \partial _2 G| \partial _1 G \rangle  
\big]
\end{eqnarray*} 
where 
$\partial_\mu =\frac {\partial  }{\partial \phi^\mu  }$, ($\mu=x,y$)
and 
where 
$|G \rangle =|G(\phi^1,\phi^2) \rangle $
is a unique gapped ground state 
of the hamiltonian, $H(\phi^1,\phi^2) $,
with
a twisted boundary condition specified by 
 $e^{i\phi^1} $ and  $e^{i\phi^2} $ 
for each  of the two translational directions
as
$H|G \rangle = E|G \rangle $.

This $C$ is the topological number (Chern number)
and is intrinsically integer as explained below. 
This topological property is clearly demonstrated by 
defining the Berry connection ( a formal form of the vector potential)
as
\begin{eqnarray*}
{\cal A } &=&  \langle G |  d G \rangle
=\langle G|\partial _\mu G \rangle d\phi^\mu 
=d\bm{\phi}\cdot \bm{A},
\qquad
\bm{A} =   \langle G|\bm{\nabla} G \rangle
\end{eqnarray*}
where $\bm{\phi}=(\phi^1,\phi^2,0) $ and $\bm{\nabla} =(\partial_1,\partial _2,\partial _3)$.
Note that, since the phase of the state $|G \rangle $ is
completely arbitrary, one may take a different choice as
$|G' \rangle =|G  \rangle \omega $, $\omega =e^{i\theta} $, 
$\theta\in \mathbb{R}$, which gives a different 
connection  ${\cal A}'=\, \langle G'| d G' \rangle 
=d \bm{\phi}\cdot \bm{A}'  $.
They are related with each other as \cite{Berry84}
\begin{eqnarray*} 
{\cal A} &=& {\cal A} '+\omega ^{-1}  d \omega 
= {\cal A} '+i d\theta,
\qquad
\bm{A}  = \bm{A} '+ i \bm{\nabla} \theta
\end{eqnarray*} 
This is the gauge transformation similar to the case of the 
electromagnetic field governed by 
the Maxwell equations. (As for the differential form, see \cite{Flanders}.) 
This is called Abelian since $\omega $ is just a complex number $|\omega|=1 $.

Since the twisted boundary condition of $e^{i2\pi}$ is equivalent to the periodic boundary
condition without any twist, the integral region 
$T^2=\{(\phi^1,\phi^2)|\phi^\mu \in [0,2\pi]\}$
is identified as the two-torus. Then the Chern number of the manybody state is written as
\begin{eqnarray*}
C= \frac {1}{2\pi i} \int_{T^2}  d \bm{S}\cdot {\rm rot}\,  \bm{A} 
= \frac {1}{2\pi i} \int_{T^2}  d{\cal A}
\equiv \frac {1}{2\pi i} \int_{T^2}  {\cal F}
= \frac {1}{2\pi i} \int_{T^2}  \langle dG | d G \rangle 
\end{eqnarray*} 
where $d \bm{S}= d\phi^1d\phi^2 (0,0,1) $ 
and the field strength is defined as ${\cal F} = d {\cal A} =\langle dG | d G \rangle $.
The connection ${\cal A}$ is gauge dependent but the Abelian field strength ${\cal F}$ 
is gauge invariant as ${\cal F}' =d {\cal A}'=d{\cal A}$ since 
$d(\omega ^{-1} d \omega )= -\omega ^{-1} d\omega  \omega ^{-1} \omega
=-( \omega ^{-1} d \omega )^2=0$ (See {\ref{app:proj_multi}).

\begin{wrapfigure}{r}{5cm}
\begin{center}
\includegraphics[width=5cm]{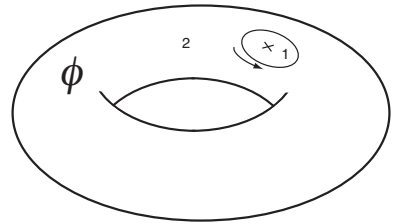}
\caption{
\small Singularity of the gauge and the two patches to avoid it. 
}
\label{fig:gauge}
\end{center}
\end{wrapfigure}
To understand the quantization of the Hall conductance, we need to specify the
gauge\cite{Kohmoto85}. Here we discuss this gauge fixing generically\cite{Hatsugai04e}.
Since the two-torus $T^2$ is boundaryless, the Chern number is always vanishing
if one is allowed to use the Stokes theorem globally 
$
C=\int_{T^2} d{\cal A} =
\int_{\partial T^2=\emptyset} {\cal A} =0
$.
To have a nonzero Chern number, the single global gauge is not allowed and 
one needs to use two different gauges, at least, say, ${\cal A}_1$ and  ${\cal A}_2$.
Let us  first
mention that the state $|G \rangle $ is gauge dependent but 
the projection into the
ground state space $P=| G \rangle \langle G |=| G' \rangle \langle G' |$ 
is gauge invariant.
Then taking arbitrary (fixed) state $|T_1 \rangle $, a normalized and
gauge fixed state $|G_1 \rangle $ is
given 
 as $|G_1 \rangle = P|T_1  \rangle/ \sqrt{N_1}$ where $N_1= \langle T_1|P|T_1 \rangle 
=|\langle G|T_1 \rangle |^2\ge 0$ is a gauge invariant normalization constant. 
It is clear that this gauge becomes singular when the normalization $N_1$ is vanishing 
at some point  in $T^2$ ($N_1=0 $).
Near this singular point $\phi_1$, one needs to use the other gauge, say $|G_2 \rangle  $
using another arbitrary state $|T_2 \rangle $ that
 is generically regular in the region $R_1$ that includes $\phi_1$, $N_2(\phi)\ne 0, \phi\in R_1$.

The gauge transformation between the two,
${\cal A}_1={\cal A}_2+i\theta_{21}
$ is given as
$|G_1 \rangle =|G_2 \rangle \omega_{21} $, $\omega_{21}=\langle T_2 | P| T_1 \rangle /\sqrt{N_1N_2} 
= e^{i\theta_{21}} $. 
When these two gauges are enough to span the whole $\phi$  space, the Chern number is
written as
\begin{eqnarray*}
C &=&\frac {1}{2\pi i}\bigg(
  \int_{R_1} d{\cal A}_2 +\int_{T^2\setminus R_1} d{\cal A}_1\bigg)
=\frac {1}{2\pi i}   \int_{\partial R_1} ({\cal A}_1-{\cal A}_2)
=\frac {1}{2\pi} \oint_{\partial R_1}d\theta_{21}
\end{eqnarray*} 
Here the reason the Chern number is integer is clear since the gauge 
transformation $\omega _{21}$ is single valued on the boundary $\partial R_1$.

When the system is non-interacting,  $V=0$, one may go further.
In this case, the ground state of the many body system, ($M$ particle state),
 $|G \rangle $ is constructed 
from  one particle states.
 It is explicitly expressed by writing 
the hamiltonian as $H = \bm{c} ^\dagger (\bm{h}) \bm{c}$
where $\bm{c} ^\dagger =(c_1,\cdots, c_N)$ and $N$ is a total number of system sites. 
The ground state is given by the filling the one particle states below the 
fermi energy $\epsilon_{F} $ as ($\epsilon_{M}\le \epsilon_{F}  $)
\begin{eqnarray*}
| G \rangle &=& |\bm{\Psi} \rangle =
\prod_{\ell=1}^M 
 \bm{c}  ^\dagger \bm{\psi} _\ell | 0 \rangle, 
\ \
\bm{h} \bm{\psi}_\ell = \epsilon_{\ell} \bm{\psi}_\ell,\quad \epsilon_{\ell}\le\epsilon_{\ell'},
(\ell<\ell'),
 \quad \bm{\psi} _\ell ^\dagger  
 \bm{\psi} _{\ell'}=\delta _{\ell\ell'}
\end{eqnarray*} 
where the many body state is labeled by the filled one particle state, 
that form a multiplet,
$\bm{\Psi}=(\bm{\psi}_1,\cdots,\bm{\psi}_M) $\cite{Hatsugai05,Hatsugai06a,Hatsugai05-char}.
The Berry connection of this non-interacting ground state has
an interesting form by the non-Abelian Berry connection $\bm{A} _M $ 
which is defined  as 
\begin{eqnarray*}
{\cal A} &=& \langle g |\sum_\ell
(\bm{c}  ^\dagger \bm{\psi} _1) \cdots
(\bm{c}  ^\dagger d\bm{\psi} _\ell) \cdots
\cdot  (\bm{c}  ^\dagger \bm{\psi} _M)
| 0 \rangle=
\sum_\ell \det_M \bm{\Psi} ^\dagger (\bm{\psi} _1,\cdots, d\bm{\psi} _\ell,\cdots, \bm{\psi} _M)
=  {\rm Tr}\, \bm{A} _M
\\
\bm{A} _M &=& 
 \bm{\Psi} ^\dagger  d \bm{\Psi} 
=\matthree
{\bm{\psi}_1 ^\dagger d \bm{\psi}_1 } {\bm{\psi}_1 ^\dagger d \bm{\psi}_2}{\cdots}
{\bm{\psi}_2 ^\dagger d \bm{\psi}_1  }{\bm{\psi}_2 ^\dagger d \bm{\psi}_2}{\cdots}
{\vdots}{\vdots}{\ddots}.
\end{eqnarray*} 
Here the choice of the one particle states below the fermi energy is 
arbitrary and one may further allow 
to mix up among them as 
\begin{eqnarray*} 
\bm{\Psi} &=& \bm{\Psi} ' \bm{\omega } ,\quad \bm{\omega } ^\dagger \bm{\omega }  =E_M
\end{eqnarray*} 
which results in the same many body state
$
| \bm{\Psi} \rangle = | \bm{\Psi} \rangle' \det \bm{\omega }
$
since $\bm{\omega }\in U(M) $ and $\det \bm{\omega }\in U(1) $.
This $\bm{A}_M $ defines a non-Abelian Berry connection. Its property 
under the gauge transformation is given as\cite{Wilczek84,Hatsugai04e}
\begin{eqnarray*}
\bm{A}_M &=&   \bm{\omega }^{-1}  \bm{A}_M' \bm{\omega} + \bm{\omega }  ^{-1} d \bm{\omega } 
\end{eqnarray*}
This is the gauge transformation of the non Abelian matrix valued 
connection spanned by the filled one particle states. 
Correspondingly the field strength (matrix) is defined as 
\begin{eqnarray*}
\bm{F} _M &\equiv & d \bm{A} _M + \bm{A} _M ^2
=\sum_{\mu <\nu }\bm{F} _{\mu \nu }d\phi^\mu d\phi^ \nu ,
\ 
\bm{F} _{\mu \nu }= \partial _\mu \bm{A}_\nu  
-\partial _\nu \bm{A}_\mu  +[\bm{A} _\mu ,\bm{A} _\nu ]
\end{eqnarray*} 
It transforms as
$
\bm{F}_M =  \bm{\omega }  ^{-1} \bm{F} _M' \bm{\omega } 
$.
This non Abelian field strength is directly related 
to
the field strength of the manybody state 
\begin{eqnarray*}
{\cal F} &=& {\rm Tr}\, \bm{F} _M
\end{eqnarray*}
which gives the TKNN formula when each Landau levels are 
well separated\cite{Thouless82}.
It transforms as
$
{\cal F} = {\rm Tr}\, \bm{F}_M=
 {\cal F}'={\rm Tr}\, \bm{F}_M'
$, which is consistent with the physical observable $\sigma _{xy}$ 
is independent of the choice of the one particle state. 

The advantage of this non-Abelian formulation is
that we do not need to sum up all Chern numbers of the filled one particle 
states. We just need to evaluated the matrix values field strength. 
This is especially useful for the graphene in a weak magnetic field. 
Since we are mainly interested in the Hall conductance
when the fermi energy is near zero. Then one need to evaluate
all Chern numbers of  Landau levels 
with negative one particle states energies (Landau levels of Dirac sea). 
The final result is expected to be of the order of unity. It implies
most of the Chern numbers of the filled Dirac sea are canceled.
That is, making one error in evaluating each 
Chern numbers causes a serious problem on the final results.

Also we mentioned here a numerical technique developed in the
lattice gauge theory to evaluate lattice topological invariants
are quite useful for the evaluation of the $U(1)$ part of the
non Abelian Berry connection, which is a two-dimensional 
analogue of the
King Smith-Vanderbilt formula for the polarization,
this is a lattice version of the one-dimensional line integral\cite{Fukui05}.
This non Abelian method is applied for calculation of
the Hall conductance of graphene
with realistic multi using 
non orthogonal bases\cite{Arai09}.

\section{Bulk edge correspondence} 
\mynote{
For bearded edges, when $\phi=0$
\begin{eqnarray*}
H &=&  -t\sum_j\bigg[
 c_\bullet ^\dagger  ( \bm{j} ) c_\circ (\bm{j} ) 
+
e^{i2\pi\phi j_1}
c_\bullet ^\dagger  (\bm{j}-\bm{e}_1   ) c_\circ (\bm{j}) 
+
 c_\bullet ^\dagger  (\bm{j} ) c_\circ (\bm{j}+\bm{e}_1-\bm{e}_2   )   + h.c.
\bigg]
\end{eqnarray*} 
Then for $\phi=0$, we have the same form Eq.\ref{eq:Dham} for the hamiltonian 
but 
$\Delta (k_1,k_2) = -t(1+e^{ik_1}+e^{i(k_1-k_2)}$
}
Based on the decades of studies for
topological states in condensed matter physics, 
it has been widely understood that 
although the bulk of topologically non trivial states
is featureless and does not show any fundamental 
symmetry breaking, 
near their boundaries and defects as local perturbation, 
there are characteristic local physics 
governed by the edge states. 
This is the {\em bulk-edge correspondence}
 that can be useful to
investigate topological phases both theoretically and 
 experimentally\cite{Hatsugai93b}.
As for the graphene, there have been two types of
interesting boundary phenomena. One is the boundary states
at the zigzag edge with/without magnetic field and 
the other is  quantum Hall edge states of the Dirac fermions.
Both of them are fundamentally controlled by the bulk
which we demonstrate here.

\begin{figure}[h]
\begin{center}
\begin{minipage}{7.5cm}
\includegraphics[width=7.0cm]{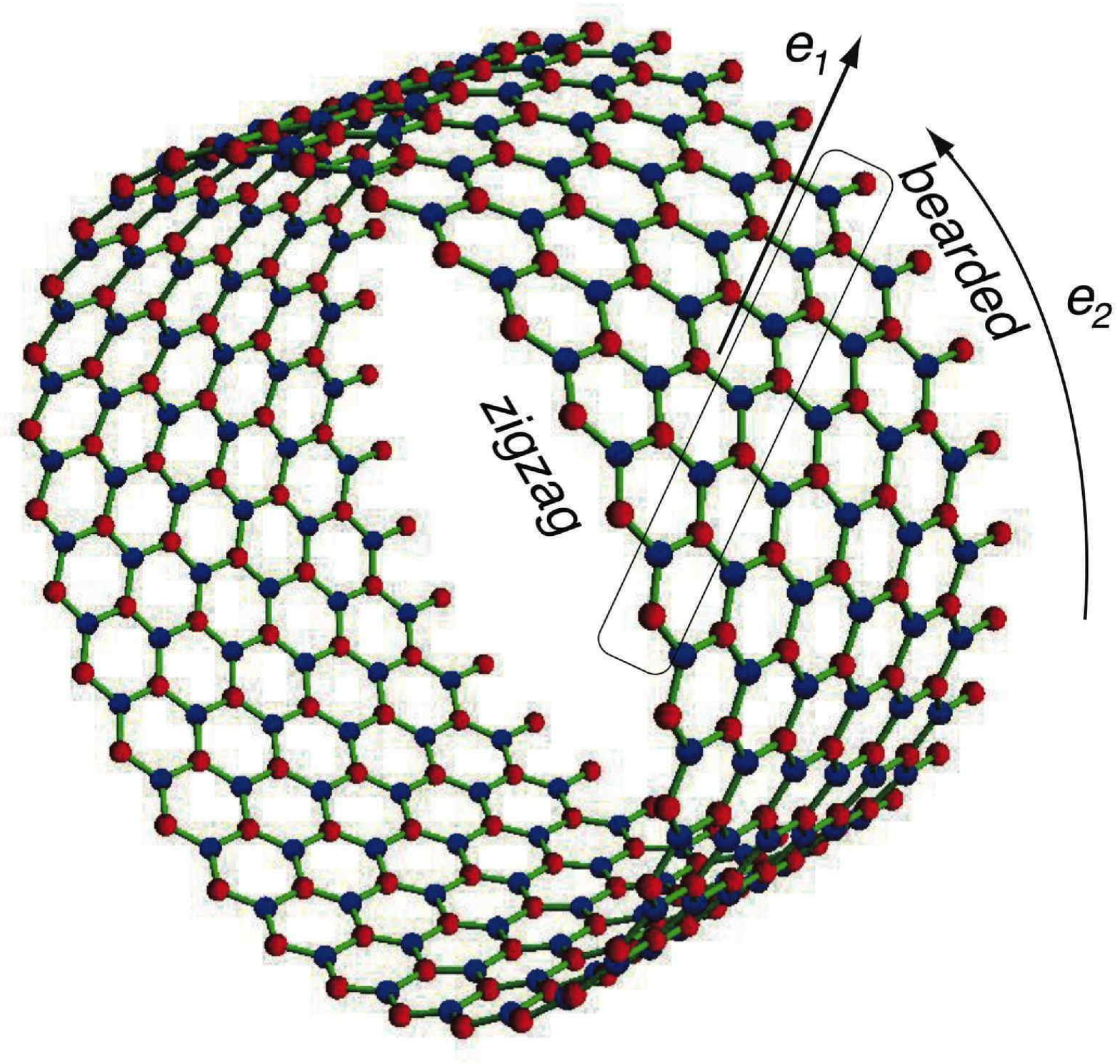}
\caption{\label{fig:cylgra}
\small Graphene on a cylinder with zigzag and bearded edges.
Unit cell of the dimensional system for the momentum representation by
$k_2$ is shown. 
}
\end{minipage}\hspace{2pc}%
\begin{minipage}{7.5cm}
\begin{center}
\includegraphics[width=5.0cm]{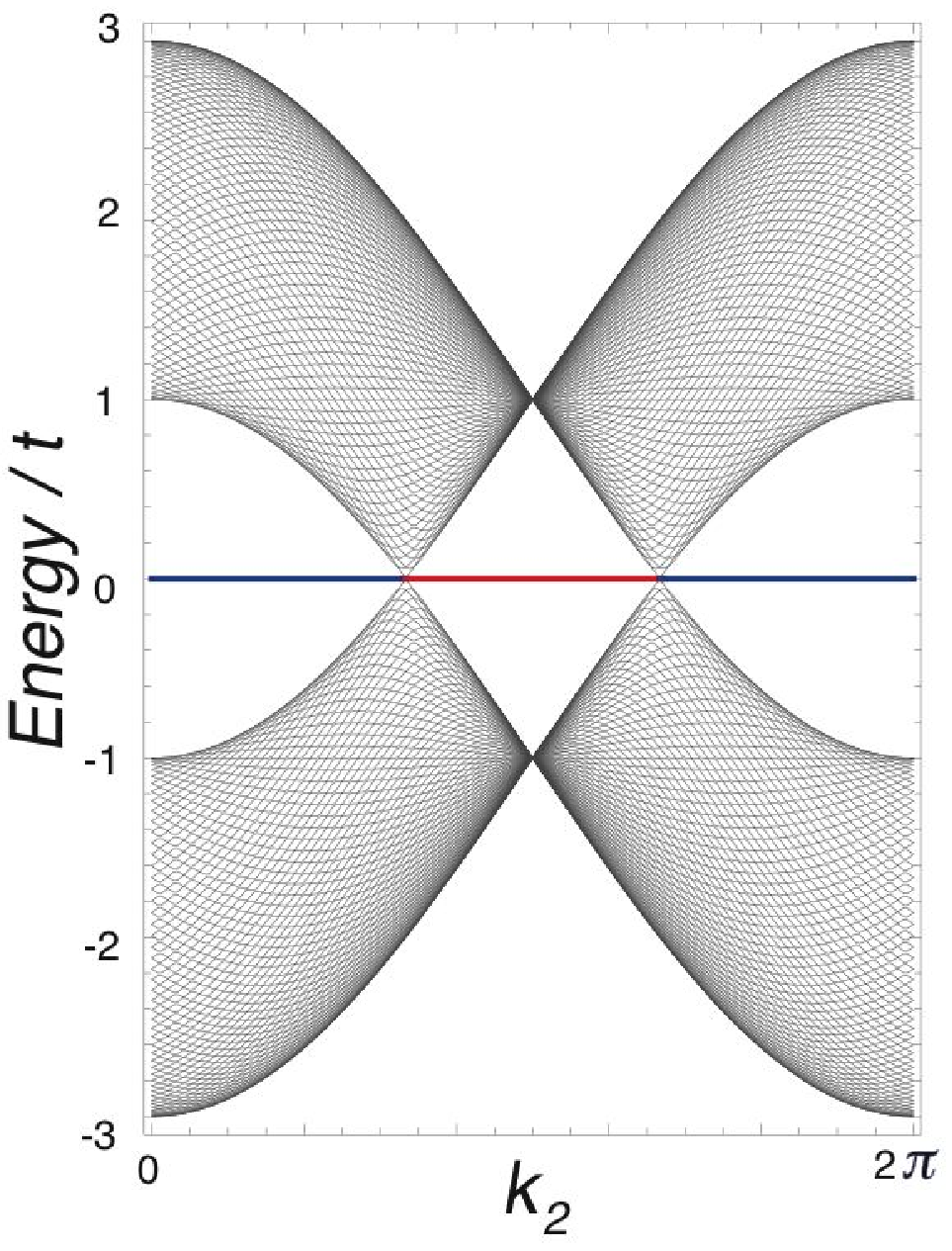}
\caption{\label{fig:zigzag}
\small One particle energy spectrum of the graphene without magnetic field
labelled by $k_2$. 
There are zero mode boundary states. The red ones are localized at
the zigzag boundary and the blue ones are localized at the bearded 
one\cite{Fujita96,Ryu02}. 
}
\end{center}
\end{minipage} 
\end{center}
\end{figure}

To investigate, let us put the graphene 
on the cylinder (Fig.\ref{fig:cylgra}) 
with zigzag-bearded edges.
Let us then take a momentum representation
only in tangential $\bm{e}_2$  direction as
$c_\alpha (\bm{j} )= \int \frac {dk_2  }{d 2\pi }
 e^{ik_2 j_2}  c_\alpha (j_1,k_2)$.
Now the total hamiltonian is decomposed 
as
$H=\int \frac {dk_2  }{d 2\pi } H _{cyl}(k_2) $
by a one dimensional hamiltonian $H_{cyl}(k_2)$ 
with 
 $k_2$
as a parameter. 
\mynote{
\begin{eqnarray*}
H = \int \frac {dk_2}{2\pi} 
H_{cyl}(k_2)
\\
c_\alpha (j) = \int \frac {dk_2}{2\pi} 
e^{ik_2j_2} c_\alpha (j_1,k_2)
\end{eqnarray*}
\begin{eqnarray*} 
H_{cyl}(k_2)
&=& \sum_{j_1}
t\bigg(
 c_\bullet ^\dagger  (j_1,k_2) c_\circ (j_1,k_2) 
+
 e^{2\pi \phi j_1 } e^{-ik_2} 
 c_\bullet ^\dagger(j_1,k_2) 
 c_\circ (j_1,k_2) 
+
 c_\bullet ^\dagger  (j_1+1,k_2) c_\circ (j_1,k_2)   
\bigg)
+ h.c.
\\
&=& \sum_{j_1}
t\bigg(
(1+ e^{i(-k_2+2\pi \phi j_1) } )
 c_\bullet ^\dagger  (j_1) c_\circ (j_1) 
+
 c_\bullet ^\dagger  (j_1+1) c_\circ (j_1)   
+ h.c.\bigg)
\\
&=& 
\big[
t_{\bullet \circ }(j_1)c_\bullet ^\dagger  (j_1+1) c_\circ (j_1)   
+
t_{\circ  \bullet }(j_1)c_\circ ^\dagger  (j_1)    c_\bullet   (j_1)  + h.c.
\big]
\\
c_\alpha (j_1) &\equiv& 
c_\alpha (j_1,k_2)
\\
t_{\bullet \circ } (j_1)&=& t
\\
t_{\circ \bullet } (j_1) &=& t(1+ e^{i(k_2-2\pi \phi j_1) } )
\end{eqnarray*} 
One particle state for each $k_2$ is expanded as
\begin{eqnarray*} 
|E \rangle &=& \sum_{j_1}
(
\psi_\bullet (j_1 ) c_\bullet ^\dagger (j_1,k_2 )
+
\psi_\circ   (j_1 ) c_\circ   ^\dagger (j_1 ,k_2)
) | 0 \rangle
\end{eqnarray*} 
}
$
H_{cyl} = 
\big[
t_{\bullet \circ }(j_1)c_\bullet ^\dagger  (j_1+1) c_\circ (j_1)   
+
t_{\circ  \bullet }(j_1)c_\circ ^\dagger  (j_1)    c_\bullet   (j_1)  + h.c.
\big]
$ 
where
$ c_\alpha (j_1) =
c_\alpha (j_1,k_2)
$,
$
t_{\bullet \circ } (j_1)= t
$ and
$
t_{\circ \bullet } (j_1) = t(1+ e^{i(k_2-2\pi \phi j_1) } )
$.
The spectrum of $H$ without magnetic field
 is shown in Fig.\ref{fig:zigzag}.
There is characteristic zero energy states which are localized near
zigzag and bearded boundaries\cite{Fujita96}. 

One particle  Schrodinger equation for each $k_2$,
$H_{cyl}(k_2)|E(k_2) \rangle =E(k_2) |E(k_2) \rangle $  by
$|E \rangle = \sum_{j_1}
(
\psi_\bullet (j_1 ) c_\bullet ^\dagger (j_1,k_2 )
+
\psi_\circ   (j_1 ) c_\circ   ^\dagger (j_1 ,k_2)
) | 0 \rangle
$
is reformulated using a transfer matrix as
($\phi=p/q$, $p$ and $q$ are mutually prime.)
\begin{eqnarray*}
\mvec
{\psi_\bullet  (q\ell+1)}
{\psi_\circ (q\ell)} 
&=& 
M^\ell
\mvec
{\psi_\bullet  (1)}
{\psi_\circ (0)} 
\end{eqnarray*} 
where $M$ is a $2\times 2 $
 matrix (see \ref{ap:tra} for the precise definition).
It implies that the boundary condition for zigzag-bearded boundaries is then 
$ \psi_\circ  (q\ell)=0$ (bearded)
and 
$\psi_\circ  (0)=0$ (zigzag).
Then the spectrum $\epsilon_\ell(k_2) $ of the edge states is determined by
$M_{21}(\epsilon_{\ell} ) = 0$ and their position is specified as 
\begin{eqnarray*} 
\text{for zigzag-bearded:     } |M_{11}(\epsilon_\ell ) | &>&  1:\quad {\rm right}\ j_1\approx \infty: {\rm (bearded\  edges)}
\\
|M_{11}(\epsilon_\ell ) | &<&  1:\quad {\rm left\ \  }\ j_1\approx 1\ : {\rm (zigzag \ edges)}
\end{eqnarray*} 
When one consider the bearded-zigzag boundaries, the 
condition is
$ \psi_\bullet   (q\ell+1)=0$ (zigzag)
and 
$\psi_\bullet  (1)=0$ (bearded). Then the spectrum is determined by
$M_{12}(\epsilon_{\ell})=0  $ and 
\begin{eqnarray*} 
\text{for bearded-zigzag:     } |M_{22}(\epsilon_\ell ) | &>&  1:\quad {\rm right}\ j_1\approx \infty: {\rm (zigzag\  edges)}
\\
|M_{22}(\epsilon_\ell ) | &<&  1:\quad {\rm left\ \  }\ j_1\approx 1\ : {\rm (bearded \ edges)}
\end{eqnarray*}

\subsection{Zero mode boundary states}
Graphene with zigzag edges without magnetic field 
has a characteristic boundary states (Fujita states)
which has been also confirmed by
a realistic first principle calculation
and also experimentally\cite{Fujita96,Okada01,Kobayashi05}.
By the general discussion above, 
we have  ($\phi=0$)  the zero modes
localized at the zigzag edge when $|M_{11}|=|1+e^{i k_2}|<1$ 
($2\pi/3 <k_2<4\pi/3$) and at
the bearded edge when $|M_{11}|=|1+e^{i k_2}|>1$ 
($0 <k_2<2\pi/3$, $4\pi/3 <k_2<2\pi$)\cite{Fujita96, Ryu02} as shown in
Fig\ref{fig:zigzag}. Actually the existence of the zero localize states is
guaranteed by the bulk\cite{Ryu02}. Let us first define the
Berry phase (Zak phase) $\gamma(k_2) $ for each $k_2$ using a bulk hamiltonian $\bm{h}(\bm{k} ) $ 
as
\begin{eqnarray*}
\gamma_{Z,B} (k_2) &=& -i \int_0^{2\pi} dk_1 \,
\bm{\psi}_{Z,B} ^\dagger \partial _{k_1} \bm{\psi}_{Z,B}=0\text{ or }\pi\ (\text{mod } 2\pi)
\end{eqnarray*} 
where $\bm{\psi}_{Z,B}$ is an eigen state of
the hamiltonian of the periodic systems $\bm{h}_{Z,B} $ (compatible with
the zigzag or bearded edges.
This quantization as $Z_2$ Berry phase is due to 
the chiral symmetry\cite{Ryu02,Hatsugai06a}. 
The zero modes which is compatible
for the unit cell of the periodic system  only exist when $\gamma(k_2)=\pi $. 
\begin{eqnarray*}
\gamma _{Z,B}(k_2) =
\mchss
{\pi\ (\text{mod } 2\pi)}
{
\text{ : zero modes exist at the zigzag (bearded) edges 
for  } k_2
}
{0\ (\text{mod } 2\pi)}
{
\text{ :  no reason to have the zero mode
for 
 } k_2
}
\end{eqnarray*} 
The condition for the zigzag boundary states, $|M_{11}|=|1+e^{-ik_2}|<1$, 
is the condition for the $\Delta_Z(\bm{k})=1+e^{-ik_2}+e^{-ik_1} $ encloses the origin
by changing $k_1:\to 2\pi$.
Since the curve is a unit circle centered at $1+e^{-ik_2}$, 
the distance between the origin and center $|1+e^{-ik_2}|$ needs to be
less than unity for the origin to be enclosed. 
{\em It establishes a bulk-edge correspondence of the zero mode boundary states for the
zigzag edges. }

With a magnetic field, there still exist zero mode boundary states though 
the $n=0$ Landau level also co-exists. Even though the bulk and the edge are both at
the same energy, the local density of state at $E=0$ shows characteristic behavior
which is consistent with the localized boundary modes\cite{Arikawa08}.

\subsection{Quantum Hall edge states of Dirac fermions}

\begin{figure}[h]
\begin{center}
\includegraphics[width=6.1cm]{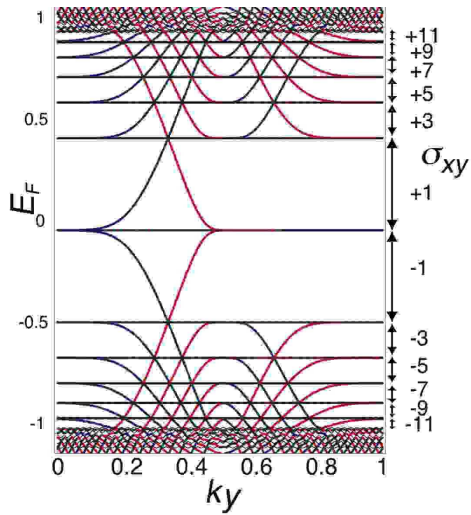}\hspace{2pc}%
\begin{minipage}[b]{20pc}
\caption{\label{fig:EdgeChern}\small
 One particle energy spectrum of the graphene on a cylinder.
The blue (red) lines 
are edge states
localized near the left (right)
edges. $k_y$ is a momentum along the edges and 
the $\sigma _{xy}$'s when the fermi energy sit there are 
shown in unit $-e^2/h$. The numbers are the Chern numbers of the bulk
and, at the same time,  numbers of the blue (red) lines in the gap 
($\phi=1/51$).
 (See also [2])
}
\end{minipage}
\end{center}
\end{figure}
Anomalous quantum Hall effect of graphene as of the Dirac fermions
is also discussed with the boundaries on the cylindrical geometry. 
Then counting the light and left edges modes between the energy gap where
the fermi energy lies, one can assign the Hall conductance using the 
Laughlin argument. Using the transfer matrix formulation given before, 
we can assign a winding number of the edge state on the complex energy surface,
which is, in our case, a Riemann surface of genus $g$ where $g$ is a number of
energy gaps (Landau gaps).
Using the setup, 
one may prove a direct relation between the two topological numbers, 
the one is the Chern number and the other is a winding number of the 
edge state 
\cite{Hatsugai93a,Hatsugai93b,Hatsugai06gra}. 
It reads physically
\begin{eqnarray*}
\sigma _{xy}^{\rm bulk} &=& 
\sigma _{xy}^{\rm edge} .
\end{eqnarray*} 
That is, the Hall conductance described by the bulk and the edges
are the same. It should be like this. 
This is the bulk-edge correspondence of the Dirac fermions
in graphene. 
In Fig.\ref{fig:EdgeChern}, energy spectrum with the zigzag-bearded boundaries
are shown as a function of $k_2$ and the corresponding Hall conductance.
It confitms the principle.

\section*{Acknowledgement}
We thank discussion with
H. Aoki, T. Kawarabayashi, T. Fukui, M. Arai, T. Morimoto and
H. Watanabe. 
The work is supported in part by Grants-in-Aid for Scientific 
Research, No.203400984 from JSPS and
No.22014002 (Novel States of Matter Induced by Frustration) 
on Priority Areas from MEXT (JAPAN).

\appendix

\section{Landau Level with anisotropic mass}
\label{sec:app1}
Let us summarize the standard Landau quantization of
electrons with parabolic dispersion with 
(effective) anisotropic mass
described by the following hamiltonian
$
\mb{H}= 
 \mb{\pi} ^\dagger  \frac {1}{2m}  {\Xi}_L   \mb{\pi}
$,
(${\rm rot}\,  \mb{A}  = B \hat z$)
where 
$\mb{\pi} = \mvec{\pi_x} {\pi_y}$,
$\pi_i = p_i-eA_i= \pi_i ^\dagger $
and 
$
\mb{\Xi}_L   
= 
\mmat
{\xi_x}
{\xi_{xy}}
{\xi_{xy}}
{\xi_y}
$ is a real symmetric anisotropy matrix.
It satisfies
\mynote{ 
$
[ \pi_x,\pi_y] = 
[ -i\hbar \partial _x-eA_x,
-i\hbar \partial _y-eA_y]
=(+i\hbar e)( \partial _x A_y-\partial _y A_x)=i\hbar e B 
$,
$
[\pi_x \frac {\ell_B}{\hbar}, 
\pi_y \frac {\ell_B}{\hbar} ]
= i$,
$
\ell_B = \sqrt{\frac {\hbar}{eB} }
$
}
$
[\pi_x ({\ell_B}/{\hbar}), 
\pi_y ( {\ell_B}/{\hbar}) ]
= i
$
 ( $\ell_B = \sqrt{\frac {\hbar}{eB} }$).
Since the matrix $\mb{\Xi}_L $ is real symmetric, it is diagonalized by
the orthogonal matrix as
$
\mb{\Xi}_L  = \mb{V} ^\dagger   \mb{\Xi}_D \mb{V} 
= {\mb{V}^\dagger  } \mb{\Xi}_D \mb{V} $
$
\mb{\Xi}_D = \text{diag}(\xi_X,\xi_Y),\ \xi_X\xi_Y=\det \mb{\Xi}_L $,
$ \xi_X+\xi_Y= {\rm Tr}\, \bm{\Xi}_L$.
$
\mb{V} = 
\mmat
{\cos\theta}{-\sin\theta}
{\sin\theta}{\cos\theta}$, $\ ^\exists\theta\in \mathbb{R}
$.
Then  we have
\mynote
{
\begin{alignat*}{1} 
H &= \mb{\Pi}  ^\dagger \mb{\Xi} _D \mb{\Pi} 
\\
\mb{\Pi}  &\equiv \mb{V} \mb{\pi} 
\\
\mvec
{\Pi_X}
{\Pi_Y}
&= \mvec
{\pi_x\cos\theta-\pi_y\sin\theta}
{\pi_x\sin\theta+\pi_y\cos\theta},
\\
[\Pi_X \frac {\ell_B}{\hbar}, 
\Pi_Y \frac {\ell_B}{\hbar} ]
&=
\big( \frac {\ell_B}{\hbar}\big)^2 
[\pi_x \cos\theta 
-\pi_y\sin\theta
,
\pi_x
 \sin\theta 
+\pi_y\cos\theta
]
\\
&= 
\big( \frac {\ell_B}{\hbar}\big)^2 
(
\cos\theta^2[\pi_x,\pi_y]
-
\sin\theta^2[\pi_y,\pi_x])=i
\end{alignat*} 
}
$
H = \mb{\Pi}  ^\dagger \mb{\Xi} _D \mb{\Pi} 
$,
$\mb{\Pi}  \equiv \mb{V} \mb{\pi} $,
$
[\Pi_X ({\ell_B}/{\hbar}), 
\Pi_Y ( {\ell_B}/{\hbar}) ]
=i$.

Now defining a bosonic operator, ($[a,a ^\dagger ] = 1$)
\mynote
{
\begin{alignat*}{1} 
a &= \frac {1}{\sqrt{2}} \frac {\ell_B}{\hbar} (\Pi_X+ i \Pi_Y)
\\
\Pi_X \frac {\ell_B}{\hbar} &= \frac {1}{\sqrt{2}} ( a+a ^\dagger )
\\
\Pi_Y \frac {\ell_B}{\hbar} &= \frac {1}{i\sqrt{2}} ( a-a ^\dagger )
\end{alignat*} 
\begin{alignat*}{1} 
\frac {1}{2i} [a+a ^\dagger , a- a ^\dagger ] &=  \frac {-1}{i}[ a, a ^\dagger ]=i
\\
[a,a ^\dagger ] &= 1
\end{alignat*} 
}
$
a =  ( {\ell_B}/{\hbar}) (\Pi_X+ i \Pi_Y)/\sqrt{2}
$,
the hamiltonian is written as
\mynote
{
\begin{alignat*}{1} 
H &= 
\frac {1}{2m} (
\xi_X \Pi_X^2+
\xi_Y \Pi_Y^2
)
\\
&= 
\frac {1}{2m} \frac {\hbar^2} {2\ell_B^2}
\bigg(
\xi_X (a+a ^\dagger )^2
-
\xi_Y (a-a ^\dagger )^2\bigg)
\\
&=  {\hbar\omega}
\frac 1
{4}
\bigg(
\xi_X (a+a ^\dagger )^2
-
\xi_Y (a-a ^\dagger )^2\bigg)
,\quad 
 {\omega} = \frac {eB}{m} 
\end{alignat*} 
}
$
H = 
\frac {\hbar\omega}
{4}
\bigg(
\xi_X (a+a ^\dagger )^2
-
\xi_Y (a-a ^\dagger )^2\bigg)
$ where
$\displaystyle{}    {\omega} = \frac {eB}{m} 
$ is a cyclotron frequency.

Now we define a new bosonic operator ($[b,b ^\dagger ]=1$)
as 
$
a = u b + v^* b ^\dagger 
$
requiring
$
[a, a ^\dagger  ] = 
[
 u b + v^* b ^\dagger ,
 u^* b ^\dagger  + v b ] = |u|^2-|v|^2=1.
$
Here noting that 
\mynote
{
\begin{alignat*}{1} 
\xi_X(a+a ^\dagger ) ^2 
-
\xi_Y(a-a ^\dagger ) ^2 
&= 
\xi_X\big\{(u+v)b+(u^*+v^*)b ^\dagger \big\}^2
-
\xi_Y\big\{(u-v)b-(u^*-v^*)b ^\dagger \big\}^2
\\
&= 
b^2
\big\{
\xi_X(u+v)^2-
\xi_Y(u-v)^2
\}
+
{b ^\dagger }^2
\big\{
\xi_X(u^*+v^*)^2-
\xi_Y(u^*-v^*)^2
\}
\\
&
+
b b ^\dagger 
\{
\xi_X |u+v|^2  
+\xi_Y |u-v|^2  
\}
+
b ^\dagger  b
\{
\xi_X |u+v|^2  
+\xi_Y |u-v|^2  
\}
\\
&= 
b^2
\big\{
\xi_X(u+v)^2-
\xi_Y(u-v)^2
\}
+
h.c.
\\
&
+
(b b ^\dagger +b ^\dagger b )
\{
\xi_X |u+v|^2  
+\xi_Y |u-v|^2  
\}
\end{alignat*} 
}
$
\xi_X(a+a ^\dagger ) ^2 
-
\xi_Y(a-a ^\dagger ) ^2 
= 
b^2
\big\{
\xi_X(u+v)^2-
\xi_Y(u-v)^2
\}
+
h.c.
+
(b b ^\dagger +b ^\dagger b )
\{
\xi_X |u+v|^2  
+\xi_Y |u-v|^2  
\}
$,
we choose as
$
\xi_X(u+v)^2 = \xi_Y (u-v)^2
$,
$
u+v =C \sqrt{ \xi_Y}
$,
$
u-v =\pm C \sqrt{ \xi_X}
$.
Assuming $\xi_X,\xi_Y>0$ and imposing 
$|u|^2-|v|^2=1$, we have $|C|^2=1/\sqrt{\xi_X\xi_Y} =
1/(\det \mb{\Xi}_L )^{1/2}
$,
$
u =\frac {\sqrt{\xi_X}+ \sqrt{\xi_Y}}{2(\det \mb{\Xi}_L )^{1/4}} 
$,
$
v =\frac {-\sqrt{\xi_X}+ \sqrt{\xi_Y}}{2(\det \mb{\Xi}_L )^{1/4}} 
$.

Finally the  hamiltonian is written as
\begin{eqnarray*} 
H &=&  \frac {1}{2} {\hbar \omega} 
(b b ^\dagger +b ^\dagger b )
|C|^2 (\xi_X\xi_Y)
= \hbar \omega _\Xi ( b ^\dagger b + \frac {1}{2} )
\end{eqnarray*}  
where 
$
\omega_\Xi=  {\omega }{\sqrt{\det \mb{\Xi}_L }} 
=\frac {eB}{m} {\sqrt{\xi_X\xi_Y}}
$.
As for the Landau degeneracy, 
we stressed that it is independent of the anisotropy
since
the magnetic length is independent of the mass.

\section{Some notations of differential form}
\label{app:proj_multi}
In this appendix, we supplement some  details for the differential forms
which may not be so popular as the standard vector analysis. 

\begin{itemize}
\item We have omitted the wedge product ($\wedge$) which is anti-commuting and
useful for the integral over the oriented surface.
For example,
$
d\phi^1\wedge d\phi^2 = -d\phi^2\wedge d\phi^1 
$ and $d\phi^1\wedge d\phi^1 = -d\phi^1\wedge d\phi^1 =0$.
It implies
\begin{eqnarray*} 
\langle d G | d G \rangle &\equiv & 
\langle d G |\wedge| d G \rangle = 
d\phi^\mu \langle  \partial _\mu   G|
\wedge
d\phi^\nu |\partial _\nu  G \rangle =
d\phi^1
\wedge
d\phi^2 \big[
\langle  \partial _1   G| \partial _2 G \rangle 
-
\langle  \partial _2  G| \partial _1 G \rangle 
\big]
\end{eqnarray*} 

\item For any  (matrix valued) one form $\mb{X} = \mb{X} _\mu  d\phi^\mu$,
 ${\rm Tr \,} \mb{X} ^2 =0$
\begin{eqnarray*} 
{\rm Tr \,} \mb{X} ^2 
={\rm Tr \,} \mb{X} _\mu   \mb{X} _\nu  d\phi^\mu  d\phi^\nu 
= 
{\rm Tr \,} \mb{X} _\nu   \mb{X} _\mu  d\phi^\mu  d\phi^\nu 
={\rm Tr \,} \mb{X} _\mu   \mb{X} _\nu  (-d\phi^\nu  d\phi^\mu )
= 
- {\rm Tr \,} \mb{X} ^2 
\end{eqnarray*} 

\item As for a matrix valued function $\bm{\omega } $,
$d \bm{\omega } ^{-1} = - \bm{\omega }^{-1} d \bm{\omega } \bm{\omega }   ^{-1} $.
It obeys from 
$0=d\bm{E}=d(\bm{\omega }\bm{\omega }^{-1} )
= (d \bm{\omega })\bm{\omega }^{-1}
+ \bm{\omega }d \bm{\omega }^{-1} $.

\item Gauge transformation of the non-Abelian gauge strength 
$ \bm{F}_M =d\bm{A}_M  + \bm{A}_M^2  $
where $\bm{A}_M = \bm{\omega }^{-1} \bm{A}_M' \bm{\omega }+ \bm{\omega } ^{-1} d \bm{\omega }      $.
\begin{eqnarray*}
\bm{F} _M 
&=& d( \bm{\omega } ^{-1} \bm{A} _M' \bm{\omega } )+
d(\bm{\omega } ^{-1} d \bm{\omega } )
+ \bm{\omega }  ^{-1} {\bm{A} _M'}^2 \bm{\omega } 
+\bm{\omega }  ^{-1} \bm{A} _M'd \bm{\omega } 
+ \bm{\omega }  ^{-1} d \bm{\omega } \bm{\omega } ^{-1} \bm{A} _M' \bm{\omega } 
+(\bm{\omega } ^{-1} d \bm{\omega }  )^2
\\
&=& 
- \bm{\omega } ^{-1}d \bm{\omega }\bm{\omega }^{-1}    \bm{A} _M' \bm{\omega } 
+ \bm{\omega } ^{-1}d \bm{A} _M' \bm{\omega } 
- \bm{\omega } ^{-1} \bm{A} _M' d\bm{\omega } 
-\bm{\omega } ^{-1}d \bm{\omega } \bm{\omega } ^{-1}   d \bm{\omega } 
\\
&&+  \bm{\omega }  ^{-1}
  {\bm{A} _M'}^2 \bm{\omega } 
+\bm{\omega }  ^{-1} \bm{A} _M'd \bm{\omega } 
+ \bm{\omega }  ^{-1} d \bm{\omega } \bm{\omega } ^{-1} \bm{A} _M' \bm{\omega } 
+(\bm{\omega } ^{-1} d \bm{\omega }  )^2
= 
\bm{\omega }  ^{-1} \bm{F} _M' \bm{\omega } 
\end{eqnarray*}

\end{itemize}

\input{trans}

\section*{References}
\bibliography{hatsugai}


\end{document}